\documentclass[prd,twocolumn,tightenlines,superscriptaddress,preprintnumbers,floatfix,nofootinbib,]{revtex4}

\usepackage{graphicx}
\usepackage{amsmath}
\usepackage{amssymb}
\usepackage{bbold}
\usepackage{color}

\newcommand{\hbx}{\hat{\bf{x}}}
\newcommand{\wt}{\widetilde}
\newcommand{\beq}{\begin{equation}}
\newcommand{\eeq}{\end{equation}}
\newcommand{\bqa}{\begin{eqnarray}}
\newcommand{\eqa}{\end{eqnarray}}

\newcommand{\dd}{ {\rm d} }
\newcommand{\6}{\partial}
\newcommand{\bn}{{\bf n}}
\newcommand{\hbn}{\hat{\bn}}
\newcommand{\heta}{\hat{\eta}}
\newcommand{\Tr}{{\rm Tr}}
\newcommand{\f}{{\rm Ag}}

\renewcommand{\Re}{{\rm Re}}

\def\Eq#1{Eq.~(\ref{#1})}

\begin{document}

\title{New algorithm for classical gauge theory simulations 
       in an expanding box}


\author{Aleksi Kurkela}
\author{Guy Moore}
\affiliation{McGill University, Department of Physics,\\
  3600 rue University, Montr\'eal QC H3A 2T8, Canada}

\date{\today}

\begin{abstract}
We propose a new algorithm for classical statistical simulations in
scalar and gauge theories undergoing a one dimensional expansion%
, which
allows simulations to study boxes of larger
transverse extent and to continue for longer times, without
losing lattice resolution in the expanding direction.
\end{abstract}

\maketitle 

\section{Introduction}

The initial condition of heavy ion collisions of asymptotically large
nuclei at asymptotically large energies is described by the color
glass condensate framework \cite{coloredglass,coloredglass2}, in which the relevant degrees of freedom right after the collision are nearly boost invariant 
large-amplitude gauge fields, or ``glasma'' fields \cite{glasma}. The subsequent evolution towards local thermal equilibrium is of 
phenomenological interest and has attracted a lot of theoretical attention. 

The large amplitudes of the fields admit a classical statistical
treatment, and the evolution of the fields at early times can 
be followed in classical statistical lattice simulations. In the case of exact boost invariance, it is enough to perform
simulations in $2+1$D
\cite{Krasnitz:1998ns,Krasnitz:2001qu,Lappi:2003bi}. However,
fluctuations are not boost
invariant, and some fluctuations are unstable to exponential
growth \cite{Romatschke:2005pm}.  Therefore the study of this system,
and particularly of the growth and fate of these fluctuations, requires
classical lattice gauge theory studies in longitudinally expanding 3+1
dimensions.
 
In the simulations, one discretizes the gauge fields, and solves the classical Yang-Mills equations for 
time evolution.  The approximate statistical boost invariance of the system makes it most convenient to 
discretize the co-moving coordinates $\tau = \sqrt{t^2-z^2}$ and $\eta={\rm atanh}(z/t)$. In these 
coordinates the physical lattice spacing in the longitudinal direction $a_\eta = \tau \Delta\eta$ grows linearly 
as a function of proper time, while the transverse lattice spacing $a_\perp$ stays
constant. This, however, introduces the bottleneck numerical challenge of these simulations:
the longitudinal spacing $\tau \Delta \eta$ cannot be much larger than
the transverse spacing even at the end of the simulation $\tau_f$; but
this requires that one start 
the simulation with an extremely fine lattice in the $\eta$-direction,
and hence with a fantastically large aspect ratio $N_\eta/N_\perp$ to have enough dynamical range in 
proper time. For a given number of lattice points 
available, this inevitably restricts simulations to boxes with small physical transverse size, making
it very difficult to accommodate all the physical scales inside the lattice. Indeed, the state 
of the art simulations in expanding lattices are being performed on
lattices with {\it e.g.},
$N_\perp^2\times N_\eta = 32^2 \times 1024$ \cite{Jurgen},
which is to be compared with simulations in static boxes reaching
$N^3=256^3$ \cite{KM3}.  This is especially problematic because the classical field
evolution becomes a multi-scale problem; there is the scale $Q_s$
associated with the structure of the initial conditions, and there is a
screening scale $m$ which at late times is parametrically
$m\sim Q_s (Q_s \tau)^{-1/2}$.  The transverse lattice spacing needs to
be fine enough to resolve the scale $Q_s$, $Q_s a_\perp < 1$, but the
transverse size should be enough to contain the scale $m$,
$m L_\perp > 1$ (with $L_\perp = N_\perp a_\perp$); otherwise important
features of the dynamics may be missed.

The purpose of this short note is to propose a new algorithm to overcome this problem, and to 
facilitate simulations with arbitrary dynamical range in time while having large $N_\perp$.
The algorithm works by cropping the lattice in half in the
$\eta$-direction
whenever the ratio $a_\eta/a_\perp\equiv \xi$ has become too large.
This procedure reduces the number
of degrees of freedom to half, which are subsequently recovered by a mesh refinement in 
the $\eta$-direction, halving the physical lattice spacing $a_\eta$. In non-abelian gauge 
theories, the mesh refinement procedure generates Gauss's law violations of the
order of $a^2 F$, which are then eliminated in the third step of the algorithm by
projecting the new field configuration to the physical manifold.

In Section \ref{sec:scalar}, we describe our algorithm in a simpler scalar theory. For the
scalar theory, the implementation is extremely simple and should be easily incorporated
into any lattice practitioner's code without effort. We then move to non-abelian gauge
theory and consider the subtleties that arise in that case. 

\section{Scalar field theory} 
\label{sec:scalar}

As a warmup, let us consider a lattice scalar
field theory discretized on a co-moving lattice with a lattice 
spacing $a_\perp$ in the transverse direction and $\Delta \eta$
in the rapidity direction. The extent of the lattice is $N_\perp^2 \times (N_\eta+1)$;
the field $\phi_{\heta, \hbn}$ and its conjugate momentum $ \pi_{\heta, \hbn}$ 
live on the lattice sites
$\hbx=(\heta,\hat{n}_1,\hat{n}_2)$, 
labeled by the indices $\heta=\{0,\ldots,N_\eta\}$ and
$\hbn=(\{0,\ldots,N_\perp-1\},\{0,\ldots,N_\perp-1\})$ in rapidity and 
transverse directions, respectively. Their evolution
in proper time is given by equations of motion
\bqa
\frac{1}{\tau} \frac{\dd}{\dd \tau} (\tau \pi_{\heta, \hbn}) 
&=& 
\left(
	\mathbf{\nabla}_\perp^2 + \frac{1}{\tau^2}\6^2_\eta
\right)
\phi_{\heta, \hbn} 
-  V'(\phi_{\heta, \hbn}),\\
\frac{\dd}{\dd \tau} \phi_{\heta, \hbn} 
&=& \pi_{\heta, \hbn},
\eqa
for a generic potential $V$.
The operators $\mathbf{\nabla}_\perp^2$ and
$\6_\eta^2$ are some implementation of lattice Laplacians, in the simplest case just symmetric differences
\bqa
\6_\eta^2  \phi_{\heta, \hbn} & = & 
\frac{1}{\Delta \eta^2}
\left(
\phi_{\heta+1, \hbn}+\phi_{\heta-1, \hbn}- 2\phi_{\heta, \hbn} 
\right),\\
\mathbf{\nabla}^2 \phi_{\heta, \hbn}&=& 
\sum_{i=1,2} \frac{1}{a_\perp^2}
\left(
\phi_{\heta, \hbn+\hat{\bf{e}}_i}+\phi_{\heta, \hbn-\hat{\bf{e}}_i}-2\phi_{\heta, \hbn}
\right),
\eqa
where the summation goes over the transverse directions.
One way to obtain this form is by deriving it from a Hamiltonian written
in terms of the lattice fields,
\bqa
\label{ScalarH} \hspace{-0.5cm}
H & = & \frac{a_\perp^2 \tau \Delta \eta}{2} \sum_{\heta,\hbn}
\bigg( 
      \pi_{\heta,\bn}^2 + V(\phi_{\heta,\hbn})
      \vphantom{\sum_{i=1,2}}
      + \frac{(\phi_{\heta+1,\hbn}-\phi_{\heta,\hbn})^2}
             {\tau^2 \Delta \eta^2} 
\nonumber \\ && \hspace{2cm}
      + \sum_{i=1,2} \frac{(\phi_{\heta,\hbn+\hat{\bf{e}}_i}
                           -\phi_{\heta,\hbn})^2}{a_\perp^2}
\bigg),
\eqa
where $a_\perp^2 \tau \Delta \eta \sum \simeq \int d^2 x_\perp \tau
d\eta$ and the three terms in parenthesis are $\pi^2$,
$\tau^{-2} (\6_\eta \phi)^2$, and $(\nabla_\perp \phi)^2$ respectively.

In practical simulations, one typically imposes periodic boundary conditions
(BC) to minimize the finite volume effects and for the transverse directions this 
is our choice. But in the rapidity direction we find it beneficial 
to impose Neumann BC, that is,
we discard the term in \Eq{ScalarH} involving
$(\phi_{N_\eta+1\equiv 0,\hbn}-\phi_{N_\eta,\hbn})^2$
or equivalently set
\bqa
\6_\eta^2  \phi_{0, \hbn} & = & 
\frac{1}{\Delta \eta^2}
\left(
\phi_{1, \hbn} - \phi_{0, \hbn} 
\right), \nonumber \\
\6_\eta^2  \phi_{N_\eta, \hbn} & = & 
\frac{1}{\Delta \eta^2}
\left(
\phi_{N_\eta-1, \hbn} - \phi_{N_\eta, \hbn} 
\right).
\label{BC}
\eqa
Like periodic BC, Neumann BC conserve the energy%
 \footnote{One could equally well use Dirichlet BC and fix the 
 field value at the boundary. However we expect $\langle \phi^2 \rangle$
 to shrink with time, so this may lead to
 larger finite size effects.}. And we will find
 that the cutting of the lattice is better implemented
 with Neumann BC.

\begin{figure}
\includegraphics*[width=0.4\textwidth]{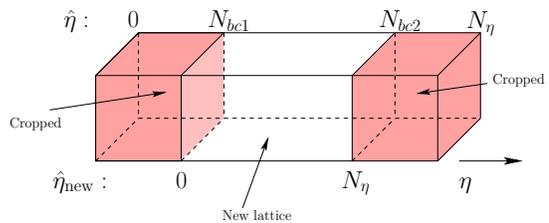}
\caption{\label{fig_latcrop}
In the first step of the algorithm, the extent of the 
lattice in the rapidity direction is reduced by a factor 2 by cropping
the ends of the lattice.  }
\end{figure}

Our algorithm works as follows: Whenever the ratio of lattice
spacings $\xi \equiv a_\eta/a_\perp$ reaches a fiducial value $\xi_c\lesssim 1$
we do the following. 
First, we divide the
lattice into three rapidity regions $\{0,\ldots, N_{bc1}-1\}$,
$\{N_{bc1}, \ldots,  N_{bc2}\}$, and $\{N_{bc2}+1 \ldots  N_\eta\}$, with
$N_{bc1} = N_\eta/4$ and $N_{bc2}= 3 N_\eta/4 $ as shown in Figure \ref{fig_latcrop}.
We discard the end caps, and
impose Neumann BC for the middle region
\bqa
\6_\eta^2  \phi_{N_{bc1}, \hbn} & = & 
\frac{1}{\Delta \eta^2}
\left(
\phi_{N_{bc1}+1, \hbn} - \phi_{N_{bc1}, \hbn} 
\right), \nonumber \\
\6_\eta^2  \phi_{N_{bc2}, \hbn} & = & 
\frac{1}{\Delta \eta^2}
\left(
\phi_{N_{bc2}-1, \hbn} - \phi_{N_{bc2}, \hbn} 
\right).
\label{BC2}
\eqa
This reduces the length of our lattice by a factor of 2.

In the second step (carried out at the same time $\tau$), we refine the mesh
in the rapidity direction. Consider
a new lattice with half the lattice spacing in the rapidity variable $a_\eta^{\rm new}=a_\eta/2$, such that
\beq
\heta_{\rm new} = 2(\heta - N_{bc1}) = \{0,\ldots, N_\eta\}.
\eeq
The original lattice field specifies the value of the new fields and momenta only
at lattice sites with even $\heta_{\rm new}$. Thus we need to provide a 
prescription how to interpolate the fields at the odd lattice sites, and the prescription 
should be such that it keeps the fields as smooth as possible, that is, it
does not transfer energy from the infrared modes to the corners of the Brillouin zone. In the 
case of scalar theory, such a description is trivial and in the simplest case
one can just linearly interpolate fields in the rapidity direction%
\footnote{Done properly, the lattice cutting and interpolation can be
  performed in the same computer memory as the fields $\phi,\pi$ are
  stored for the evolution.}
\beq
\phi_{\heta_{\rm{new}}, \hbn} 
= \frac{1}{2}\left( \phi_{\heta_{\rm{new}}-1, \hbn} 
+ \phi_{\heta_{\rm{new}}+1, \hbn}\right), \textrm{  for odd $\heta_{\rm{new}}$}.
\eeq
In terms of the new $\eta$-variable, the Neumann BC reads as in \Eq{BC}.
This completes the algorithm. 

Without using our algorithm, there are lattice spacing errors which
scale as ${\mathcal O}(a_\perp^2 k_\perp^2)$ and as
${\mathcal O}(a_\eta^2 k_\eta^2) 
 = {\mathcal O}(\Delta \eta^2 k_\eta^2 \tau^2)$.
The latter error grows with time.  By refining the lattice, we prevent
this growth with time.
However, there are three places where new systematic errors arise,
though each proves to be manageable.
Firstly, using Neumann BC
physically corresponds to replacing the end of the lattice with a
mirror, leading to unphysical interference effects within one 
correlation length ($\Delta (\eta \tau) \sim 1/k_\eta$) of the boundary. 
This can be easily ameliorated by performing all measurements only in a
fiducial volume away from the boundary.  Very conservatively, one may
define expectation values by
\beq
\langle \mathcal{A} \rangle \equiv \sum_{\hbn}\sum_{\heta 
= N_\eta/4}^{3/4 N_\eta} \mathcal{A}_{\heta, \hbn},
\eeq
and use lattices with aspect ratio $\chi \equiv N_\eta / N_\perp \geq 2$.

Secondly, abruptly imposing Neumann BC creates a configuration in 
the new lattice which has a cusp in the fields where the new boundary
lies. The cusp contains unphysical ultraviolet modes which subsequently 
propagate to  the region where measurements are performed and thus 
cause contamination. However, one can introduce the new BC
adiabatically as follows.  Rather than abruptly
introducing ``mirrors'' at $N_{bc1}$ and $N_{bc2}$, one gradually
introduces ``partly silvered'' mirrors, with silvering fraction $\f$.
Specifically, one multiplies the
$(\phi_{N_{bc2}+1,\hbn}-\phi_{N_{bc2},\hbn})^2$
and $(\phi_{N_{bc1},\hbn}-\phi_{N_{bc1}-1,\hbn})^2$
terms in \Eq{ScalarH} by a coefficient $[1-\f(\tau)]$,
which smoothly changes from $\f(\tau)=0$ a few correlation times before
the cut is to be made, to $\f(\tau)=1$ at the time when the cut is to be made.
When $\f(\tau)=0$ (zero silvering) the lattice obeys the usual update rules.
When $\f(\tau)=1$ (full silvering), there are Neumann BC's at $N_{bc1}$
and $N_{bc2}$.  For
intermediate values, waves propagating towards the $N_{bc1}$ plane from
either side will
partially reflect as from a half-silvered mirror, with $\f(\tau)$ the
extent of ``silvering'' of the mirror.  Once $\f(\tau)=1$,
equations of motion for points
between $\heta = N_{bc1}$ and $\heta=N_{bc2}$ inclusive no longer
make any reference to points outside this range, so the cut is then
harmless.

Thirdly, in performing the interpolation, one systematically 
underestimates the energy density by an $\mathcal{O}(a_\eta^2 k_\eta^2)$
amount. This is, however, a subleading source of error in
the simulation compared to the discretization error from 
the transverse directions $\mathcal{O}(a_\perp^2 k_\perp^2)$ for two reasons. First,
as long as one keeps the fiducial ratio $\xi_c<1$, the lattice spacing in the 
$\eta$ direction is smaller that the transverse one. And secondly, due 
to the longitudinal expansion, $k_\eta$ gets redshifted and one generally expects
$k_\eta < k_\perp$. Hence, during the whole simulation, up to arbitrarily late times,
the systematical errors are bounded by those arising from the discretization of the
transverse direction. 

\section{Gauge theory}

Now we apply these ideas to
nonabelian gauge theory defined on the same anisotropic
expanding lattice. The Hamiltonian  (in $A_\tau=0$ gauge) is%
\footnote{%
    Our normalization is related to that of \cite{Jurgen} by 
    $E^{\rm here}_\alpha = c_\alpha/a_\tau E_\alpha$ in the
    notation of the reference. The usual continuum normalization of
    electric fields is $E_\alpha =a_\alpha E^{\rm cont}_\alpha$. }
\bqa
H &=& a_\perp^2 a_\eta \sum_{\heta,\hbn} \Big\{  \sum_{i=\eta,1,2}\Tr
\big[a_i^{-2} E^{i\,2}_{\heta,\hbn}  \big]  \nonumber \\
&&+ \frac{2}{a_\perp^2 a_\eta^2 }\sum_{i=1,2} \Re \Tr[ \mathbb{1} -
  \underline\square^{i,\eta}_{\heta,\hbn} ] \nonumber \\
&&  + \frac{2}{a_\perp^4} \Re \Tr[ \mathbb{1} -
  \underline\square^{1,2}_{\heta,\hbn} ] \Big\},
\eqa
where the plaquette operator $\underline\square$ is the product of 
the link matrices around an elementary plaquette
\bqa
\underline\square^{i,j}_{\hbx} = 
   U^{i}_{\hbx} U^{j}_{\hbx+\hat{\bf{e}}_i}
   U^{i\,\dagger}_{\hbx+\hat{\bf{e}}_j} U^{j\,\dagger}_{\hbx},
\eqa
where the link matrices $U_{\heta, \hbn}$ are elements of the group,
and their canonical momenta $E_{\heta, \hbn}$ belong to the Lie 
algebra of the group. 
The corresponding equations of motion read
\bqa
\label{Uem}
\frac{\dd}{\dd \tau} U^i_{\heta, \hbn} &=& i E^i_{\heta, \hbn} U^i_{\heta, \hbn}, \\
\frac{\dd}{\dd \tau} \: \frac{a_\eta}{a_i^2} E^i_{\heta, \hbn} &=& 
{\rm ad}\left[ \sum_{j\neq i} \frac{-ia_\eta}{a_i^2 a_j^2}
\left( \underline\square^{i,j}_{\heta,\hbn} + \overline{\square}^{i,j}_{\heta,\hbn} \right)
\right],
\label{Eem}
\eqa
where ${\rm ad}[U]^a = 2 \Re\,\Tr \:t^a U$ denotes the adjoint representation, where 
$t^a$ are the generators of the lie algebra in canonical normalization
$\Tr t^a t^b = \frac{1}{2} \delta^{ab}$, and
\beq
\overline\square^{ij}_{\hbx} \equiv
   U^{i}_{\hbx} U^{j\,\dagger}_{\hbx+\hat{\bf{e}}_i-\hat{\bf{e}}_j}
   U^{i\,\dagger}_{\hbx-\hat{\bf{e}}_j}U^{j}_{\hbx-\hat{\bf{e}}_j}
\,.
\eeq

The Neumann BC is implemented by setting the links and electric
fluxes penetrating
the boundary to zero
\bqa
U_{-1,\hbn}^{\eta}=0 , \quad
E_{-1,\hbn}^\eta = 0 , \quad
U_{N_\eta,\hbn}^{\eta}=0  , \quad
E_{N_\eta,\hbn}^\eta = 0  ,
\eqa
or equivalently by discarding the terms including these links
from the Hamiltonian. 

In analogy to the scalar field theory, the first step
is to cut the lattice  in half adiabatically by modifying the Hamiltonian.
This is done by multiplying the terms in Hamiltonian containing 
plaquettes connecting the different rapidity regions 
$(\underline\square^{i,\eta}_{N_{bc1}-1,\hbn}$ and $\underline\square^{i,\eta}_{N_{bc2},\hbn})$ by the 
silvering function $[1-\f(\tau)]$
which again smoothly goes from $\f(\tau)=0$ to $\f(\tau)=1$. This modification 
makes sure that the resulting configuration, when the mirror becomes
fully reflecting, is smooth.

An extra complication that arises in the gauge theory comes from the need
to satisfy Gauss' law
\bqa
\label{Gausslaw}
\sum_{\alpha=1,2,\eta}
\frac{1}{a_\alpha^2}\left(
E^\alpha_{\hbx} 
- U^{\alpha\,\dagger}_{\hbx-\hat{e}_\alpha} E^\alpha_{\hbx-\hat{e}_\alpha}U^{\alpha}_{\hbx-\hat{e}_\alpha}
\right) = 0 \,.
\eqa
Before cropping the lattice, Gauss' law at the site $(N_{bc2},\hbn)$
receives a contribution from $E^\eta_{N_{bc2},\hbn}$.  Abruptly
discarding everything to the right of $N_{bc2}$ will discard this flux,
and Gauss' law will not be satisfied on the $N_{bc2}$ surface (or the
$N_{bc1}$ surface).  Physically this means that there will be
``charges'' trapped on the surface, representing the flux which
previously propagated into the mirror.
The flux can, however, be forced
to go to zero by also adiabatically removing the terms in the Hamiltonian containing
color-electric fields penetrating the mirrors, \emph{i.e.}, by multiplying the terms 
containing $(E^{\eta}_{N_{bc1}-1,\hbn})^2$ and $(E^{\eta}_{N_{bc2},\hbn})^2$ by the
function $[1-\f(\tau)]$.  The contribution of these electric fields to
\Eq{Gausslaw} get multiplied by $[1-\f(\tau)]$, while \Eq{Eem} changes
from involving $\frac{\dd}{\dd\tau} E^\eta_{N_{bc2},\hbn}/a_\eta$
to involving  $\frac{\dd}{\dd\tau} [1-\f(\tau)]
E^\eta_{N_{bc2},\hbn}/a_\eta$, which makes the $E$ field grow
correspondingly, preserving Gauss' law.  But wave evolution will mean
that the $E$ field naturally evolves to remain about the same size, so
its contribution to Gauss' law will shrink to zero as $\f(\tau)$
approaches 1.

\begin{figure}
\includegraphics*[width=0.4\textwidth]{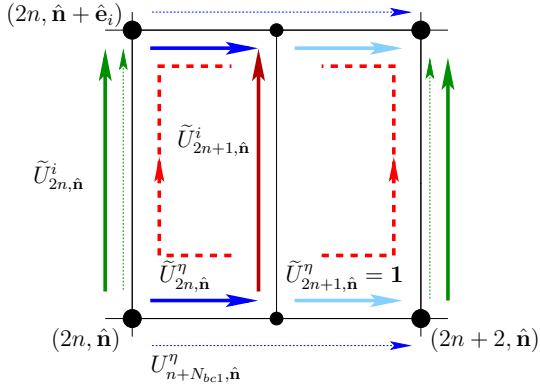}
\caption{\label{fig_interpolate}
The interpolation of the link matrices.
The thin dotted arrows correspond to the links of the original lattice,
while the thick arrows denote link matrices on the new finer lattice. 
The the color coding (if available)
indicates the assignments of the old matrices to the new ones.
Links in the $\eta$-direction starting from odd $\eta$-sites on the
the new lattice are set to unit matrices (light blue arrows) while the transverse
links starting from odd $\eta$-sites are unitarized averages
of the parallel transports along the paths shown by the dashed
lines.
}
\end{figure}

In the second step of the algorithm, the mesh in the rapidity direction is
refined and fields are interpolated. We begin by
describing the procedure to interpolate the link matrices followed by
the electric fields.  The process of interpolating the gauge field links
is illustrated in Figure \ref{fig_interpolate}.  The transverse links
between previously existing lattice sites are left unchanged; that is,
we choose $\wt{U}^i_{2n,\hbn} = U^i_{n+N_{bc1},\hbn}$ for $i=1,2$,
where we denote the fields on the refined lattice with tildes, and give 
their coordinates in terms of the refined rapidity variable $\hat{\eta}_{\rm new}=2(\heta-N_{bc1})$.
Next consider the two sites $(n+N_{bc1},\hbn)$ and $(n+N_{bc1}+1,\hbn)$
on the old lattice, which become $(2n,\hbn)$ and $(2 n+2,\hbn)$ on the refined lattice.  
The new $\eta$-links should obey
\beq
\wt{U}^\eta_{2n,\hbn} \wt{U}^\eta_{2n+1,\hbn} = U^\eta_{n+N_{bc1},\hbn}
\eeq
so that comparisons between the sites on the refined lattice agree with
those on the unrefined lattice.  There is a new gauge freedom on the
introduced point $(2n+1,\hbn)$, and we will use up that freedom to
choose $\wt{U}^\eta_{2n+1,\hbn}={\mathbf 1}$ the identity.  Then
$\wt{U}^\eta_{2n,\hbn} = U^\eta_{n+N_{bc1},\hbn}$.

Next we interpolate the transverse links of form $\wt{U}^{i}_{2n+1,\hbn}$,
$i=1,2$.  We take the connection between the site $(2n+1,\hbn)$ and the
site $(2n+1,\hbn+\hat{\bf{e}}_i)$ to be the re-unitarized average of the
connections along the two closest paths which use the $\eta$-links and
the pre-existing transverse links, as shown in Figure
\ref{fig_interpolate}:
\bqa
\wt{U}^i_{2n+1,\hbn} & = & {\rm Proj}_{{\rm SU}(N)} \frac{1}{2} \left(
 \wt{U}^\eta_{2n+1,\hbn} \wt{U}^i_{2n+2,\hbn} 
 \wt{U}^{\eta\,\dagger}_{2n+1,\hbn+\hat{\bf{e}}_i}
\right. \nonumber \\ && \left. \hspace{1.7cm}
+\wt{U}^{\eta\,\dagger}_{2n,\hbn} \wt{U}^i_{2n,\hbn}
 \wt{U}^\eta_{2n,\hbn+\hat{\bf{e}}_i} \right),
\eqa
where ${\rm Proj}_{{\rm SU}(N)}$ means projection onto the nearest
group element, which is just a rescaling in SU(2) and can be performed
for ${\rm SU}(N>2)$ using the algorithm of Ref.~\cite{Liang:1992cz}.

With the links defined, we turn to the electric fields.  Though \Eq{Eem}
defines the electric field $E^\alpha_{\hat\eta,\hbn}$ as an adjoint object transforming at the
site $(\hat\eta,\hbn)$, the $E^\alpha$-fields are
conjugate variables of the link variables $U^\alpha$ and hence it is
most natural to think of them as ``living'' on the
links; the natural objects are in fact $E^\alpha U^\alpha$ which
belong to the tangent space of $U^\alpha$ rather than the lie algebra 
elements $E^\alpha$.

We choose the $E^i$-fields on the existing transverse links to be unchanged,
$\wt{E}^i_{2n,\hbn}=E^i_{n+N_{bc1},\hbn}$.
The $E^\eta$-fields on the refined lattice, viewed as living at the
centerpoints of their links, lie $\frac{1}{4}$ of the way between
centerpoints of the links on the unrefined lattice, and should therefore
be interpolated using
\bqa
8\wt{E}^{\eta}_{2n,\hbn} & = & 3 E^{\eta}_{n+N_{bc1},\hbn}
                     + U^{\eta\,\dagger}_{n+N_{bc1}-1,\hbn} \times
\nonumber \\ && {} \times
                       E^{\eta}_{n+N_{bc1}-1,\hbn}
                       U^\eta_{n+N_{bc1}-1,\hbn} \,,
\nonumber \\
8\wt{E}^{\eta}_{2n+1,\hbn} & = & 
 3 \wt{U}^{\eta\,\dagger}_{2n,\hbn} E^{\eta}_{n+N_{bc1},\hbn}
   \wt{U}^{\eta}_{2n,\hbn}
\nonumber \\ && {}
 + \wt{U}^{\eta}_{2n+1,\hbn} E^\eta_{n+N_{bc1}+1,\hbn}
   \wt{U}^{\eta\,\dagger}_{2n+1,\hbn} \,,
\eqa
where the factor on the LHS is 8 (rather than 4) because the
lattice $E^\alpha$ is $a_\alpha$ times the continuum $E^\alpha$, and
$a_\eta$ is being
reduced by a factor of 2.  This choice ensures that Gauss' law,
\Eq{Gausslaw}, remains identically satisfied at even-$\hat\eta_{\rm new}$ lattice points
$(2n,\hbn)$.

\begin{figure}
\includegraphics*[width=0.33\textwidth]{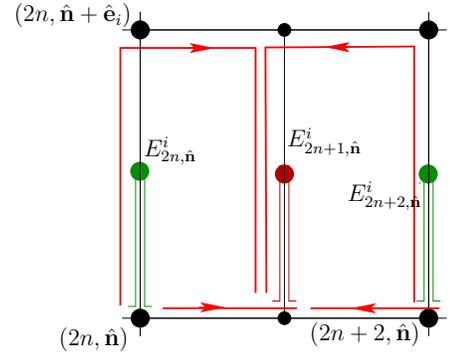}
\caption{\label{fig_einterpolate}
Interpolation of the transverse electric fields emanating from
the new lattice sites. The graphical representation
emphasizes that while under gauge transformations the $E$-fields transform as objects ``living'' on
sites, they are conjugate variables of the
gauge links and hence live between sites. The transverse $E$-field on the new lattice site is the average 
of electric fields on neighboring sites, parallel transported 
along the paths shown by the arrows. The fundamental
(anti-fundamental) indices are parallel transported along the
short (long) paths. }
\end{figure}

We define the transverse electric fields emanating from the new sites as
the average of the parallel transports of the electric fields $\pm 1$
units of $\eta$ away, as shown in Figure \ref{fig_einterpolate}.  We
parallel transport the indices of
$E^i U^i$ in the same way as we did with the link matrix $U^i$, leading
to
\bqa
\wt{E}^i_{2n+1,\hbn} & \!=\! &\nonumber\\
\frac{1}{2} {\rm ad} & \!\Bigg[\! &  
\wt U^\eta_{2n+1,\hbn} \wt E^i_{2n+2,\hbn} \wt U^i_{2n+2,\hbn} 
  \wt U^{\eta\,\dagger}_{2n+1,\hbn+\hat{\bf{e}}_i} 
  \wt U^{i\,\dagger}_{2n+1,\hbn} \nonumber \\ 
&& {}
+ \wt U^{\eta\,\dagger}_{2n,\hbn} \wt E^i_{2n,\hbn} \wt U^i_{2n,\hbn}
  \wt U^\eta_{2n,\hbn+\hat{\bf{e}}_i} \wt U^{i,\dagger}_{2n+1,\hbn} \Bigg] \,.
\eqa
Together with our choice of $\wt E^\eta$ fields, this ensures
that Gauss' law is satisfied also at the new (odd-$\hat\eta_{\rm new}$) lattice
sites --- up to corrections which arise due to the failure of $E$ to
commute with the magnetic field, which first arise at order
$a_\perp^2 a_\eta^2 EB^2$.  Therefore in the nonabelian context Gauss'
law will {\sl not} be identically satisfied at the new interpolated
lattice sites, though the failure is small and suppressed by
$a_\perp^2 a_\eta^2$.  A final step is needed, in which one corrects the
electric fields such that Gauss' law becomes exact.  The optimal choice
is to change the electric fields by an amount which is strictly the
gradient of a scalar potential, and with this choice the shift in the
electric fields, to restore Gauss' law, is unique.  We propose to use
the algorithm for finding this shift, described in detail in
Ref.~\cite{Moore:1996qs}.  This completes the specification of our
algorithm for the nonabelian case.

\section{Discussion}
We have presented a new approach to studying classical field theory in a
linearly expanding system on the lattice, with intended applications in
the study of early-time dynamics after heavy ion collisions.  The
algorithm allows the evolution to proceed to late times in boxes of
large transverse extent, without encountering the problem that the
longitudinal ($\eta$) lattice spacing becomes coarse.  We do this by
periodically cropping the box in the $\eta$ direction and refining the
mesh.  We have presented a detailed algorithm both for scalars and for
nonabelian gauge fields.

In practical applications it will probably be necessary to start the
evolution at very early times, perhaps even $\tau \sim a_\perp$ but in
any case $\tau \ll L_\perp$.  We do not think it necessary to begin with
a range of $\eta$ larger than a few, since few excitations will ever
propagate over a range of $\eta$ larger than 1 or 2.  Therefore it might
make sense to begin with a lattice which {\sl does} have a large
hierarchy $a_\eta \ll a_\perp$.  In this case our cropping and
interpolation would only begin when $a_\eta \sim a_\perp$, at times
$\tau \sim L_\perp$.

We emphasize again the importance of establishing that simulations,
particularly of nonabelian gauge fields, are really in the large
$L_\perp$ limit.  This limit is challenging, because as we emphasized in
the introduction, the screening length scale grows as $\tau^{1/2}$.
Therefore we expect that our approach will actually be necessary to
study sufficiently wide boxes, particularly if the study is to proceed
to late times.

\section*{Acknowledgments}

We thank J\"urgen Berges and S\"oren Schlichting for useful discussions
and giving us an advanced copy of Ref.~\cite{Jurgen}, which inspired
this work.  This work was supported in part by the Institute for
Particle Physics (Canada) and the National Science and Engineering Research
Council (NSERC) of Canada.


\begin{thebibliography}{99}



\bibitem{coloredglass}
For reviews see for instance,
E.~Iancu, R.~Venugopalan,
  In *Hwa, R.C. (ed.) et al.: Quark gluon plasma* 249-363.
  [hep-ph/0303204];
F.~Gelis, E.~Iancu, J.~Jalilian-Marian and R.~Venugopalan,
  Ann.\ Rev.\ Nucl.\ Part.\ Sci.\  {\bf 60}, 463 (2010)
  [arXiv:1002.0333 [hep-ph]].

\bibitem{coloredglass2}
L.~D.~McLerran, R.~Venugopalan,
  Phys.\ Rev.\  {\bf D49}, 2233-2241 (1994).
  [arXiv:hep-ph/9309289 [hep-ph]];
  Phys.\ Rev.\  {\bf D49}, 3352-3355 (1994).
  [hep-ph/9311205].
%
J.~Jalilian-Marian, A.~Kovner, L.~D.~McLerran, H.~Weigert,
  Phys.\ Rev.\  {\bf D55}, 5414-5428 (1997).
  [hep-ph/9606337].

\bibitem{glasma}
A.~Kovner, L.~D.~McLerran, H.~Weigert,
  Phys.\ Rev.\  {\bf D52}, 6231-6237 (1995).
  [hep-ph/9502289].
A.~Krasnitz, R.~Venugopalan,
  Phys.\ Rev.\ Lett.\  {\bf 84}, 4309-4312 (2000).
  [hep-ph/9909203].
T.~Lappi and L.~McLerran,
  Nucl.\ Phys.\  A {\bf 772}, 200 (2006)
  [arXiv:hep-ph/0602189].

\bibitem{Krasnitz:1998ns} 
  A.~Krasnitz and R.~Venugopalan,
  Nucl.\ Phys.\ B {\bf 557}, 237 (1999)
  [hep-ph/9809433].

\bibitem{Krasnitz:2001qu} 
  A.~Krasnitz, Y.~Nara and R.~Venugopalan,
  Phys.\ Rev.\ Lett.\  {\bf 87}, 192302 (2001)
  [hep-ph/0108092].

\bibitem{Lappi:2003bi}
  T.~Lappi,
  Phys.\ Rev.\ C {\bf 67} (2003) 054903
  [hep-ph/0303076].

\bibitem{Romatschke:2005pm} 
  P.~Romatschke and R.~Venugopalan,
  Phys.\ Rev.\ Lett.\  {\bf 96}, 062302 (2006)
  [hep-ph/0510121].

\bibitem{Jurgen} 
  J.~Berges and S.~Schlichting,
  arXiv:1209.0817 [hep-ph].
  
\bibitem{KM3}
  A.~Kurkela and G.~D.~Moore,
  arXiv:1207.1663 [hep-ph].

\bibitem{Liang:1992cz} 
  Y.~Liang, K.~-F.~Liu, B.~-A.~Li, S.~J.~Dong and K.~Ishikawa,
  Phys.\ Lett.\ B {\bf 307}, 375 (1993)
  [hep-lat/9304011].

\bibitem{Moore:1996qs} 
  G.~D.~Moore,
  Nucl.\ Phys.\ B {\bf 480}, 657 (1996)
  [hep-ph/9603384].

\end{thebibliography}
\end{document}